\documentclass[twocolumn]{aastex62}

\usepackage{CJK}
\usepackage{txfonts}
\usepackage{graphicx}
\usepackage{graphicx}
\usepackage{soul}

\usepackage{natbib}
\usepackage{natbib}

\begin{document}
\title{Spectropolarimetric inversions of the C\lowercase{a} \lowercase{{\sc{ii}}} 8542 \AA\  line in a M-class solar flare}

\correspondingauthor{D. Kuridze}
\email{dak21@aber.ac.uk}

\author{D. Kuridze}
\affil{Institute of Mathematics, Physics and Computer Science, Aberystwyth University, Ceredigion, Cymru, SY23 3Ä, UK}
\affil{Astrophysics Research Centre, School of Mathematics and Physics, Queen's University Belfast, Belfast BT7~1NN, UK}
\affil{Abastumani Astrophysical Observatory at Ilia State University, 3/5 Cholokashvili avenue, 0162 Tbilisi, Georgia}

\author{V. M. J. Henriques}
\affiliation{Rosseland Centre for Solar Physics, University of Oslo, P.O. Box 1029 Blindern, NO-0315 Oslo, Norway}
\affiliation{Institute of Theoretical Astrophysics, University of Oslo, P.O. Box 1029 Blindern, NO-0315 Oslo, Norway}
\affil{Astrophysics Research Centre, School of Mathematics and Physics, Queen's University Belfast, Belfast BT7~1NN, UK}

\author{M. Mathioudakis}
\affiliation{Astrophysics Research Centre, School of Mathematics and Physics, Queen's University Belfast, Belfast BT7~1NN, UK}

\author{L. Rouppe van der Voort}
\affiliation{Rosseland Centre for Solar Physics, University of Oslo, P.O. Box 1029 Blindern, NO-0315 Oslo, Norway}
\affiliation{Institute of Theoretical Astrophysics, University of Oslo, P.O. Box 1029 Blindern, NO-0315 Oslo, Norway}

\author{J. de la Cruz Rodr\'{\i}guez}
\affiliation{Institute for Solar Physics, Department of Astronomy, Stockholm University, AlbaNova University Centre, SE-106 91 Stockholm, Sweden}

\author{M. Carlsson}
\affiliation{Rosseland Centre for Solar Physics, University of Oslo, P.O. Box 1029 Blindern, NO-0315 Oslo, Norway}
\affiliation{Institute of Theoretical Astrophysics, University of Oslo, P.O. Box 1029 Blindern, NO-0315 Oslo, Norway}

\date{received / accepted }

\begin{abstract}

We study the M1.9 class solar flare SOL2015-09-27T10:40 UT using high-resolution full-Stokes imaging spectropolarimetry  
of the Ca {\sc{ii}} 8542 {\AA} line obtained with the CRISP imaging spectropolarimeter at the Swedish 1-m Solar Telescope. 
Spectropolarimetric inversions using the non-LTE code NICOLE are used to construct semi-empirical models 
of the flaring atmosphere to investigate the structure and evolution of the flare temperature 
and magnetic field. A comparison of the temperature stratification in 
flaring and non-flaring areas reveals strong heating of the flare ribbon during the flare peak. 
The polarization signals of the ribbon in the chromosphere during the flare maximum become stronger 
when compared to its surroundings and to pre- and post- flare profiles. 
Furthermore, a comparison of the response functions to perturbations in the line-of-sight magnetic field and temperature in 
flaring and non-flaring atmospheres shows that during the flare the Ca {\sc{ii}} 8542 {\AA} line is more sensitive to the 
lower atmosphere where the magnetic field is expected to be stronger.   
The chromospheric magnetic field was also determined with the weak-field approximation which led to results similar to those obtained with the NICOLE inversions.

\end{abstract}

%%%%%%%%%%%%%%%%%%%%%%%%%%%%%%%%%%%%%%%%%%%%%%%%%%%%%%%%%%%%%%%%%%%%%%%%%

\section{Introduction}

Solar flares are sudden energy releases within active regions caused by the reconfiguration of the coronal magnetic field resulting to temperatures as high as 20 MK. A significant amount of the released energy is 
transported along the magnetic loops to the lower layers of the solar atmosphere
via accelerated particles, magneto-hydrodynamic waves, radiation, and thermal conduction \citep{1974SoPh...34..323H}.
The vast majority of the flare energy is dissipating after reaching the dense footpoints of the magnetic loops. As a result
the majority of the flare radiative losses originates in the chromosphere and photosphere \citep{2011SSRv..159...19F}.
The lower solar atmosphere is therefore key to our understanding of the physics of solar flares. 

Due to the high complexity of the flare phenomenon, advanced, high-resolution spectropolarimetric observations  
are required to investigate the thermodynamic properties of the atmosphere including the characteristics of the magnetic field in the flare chromosphere. High-resolution, full Stokes chromospheric spectropolarimetry from ground-based telescopes offer the opportunity to obtain such datasets. However, the short duration, small spatial characteristics and unpredictable nature of flare events,  makes it difficult to capture them with the small field-of-view (FOV) of large aperture ground-based telescopes.
Furthermore, the weak polarization signals, limited number of chromospheric spectral lines suitable for magnetic field diagnostics and  non-LTE effects, makes the spectropolarimetric observations of the flare chromosphere difficult to interpret.

\begin{figure*}[t]
\begin{center}
\includegraphics[width=18.1cm]{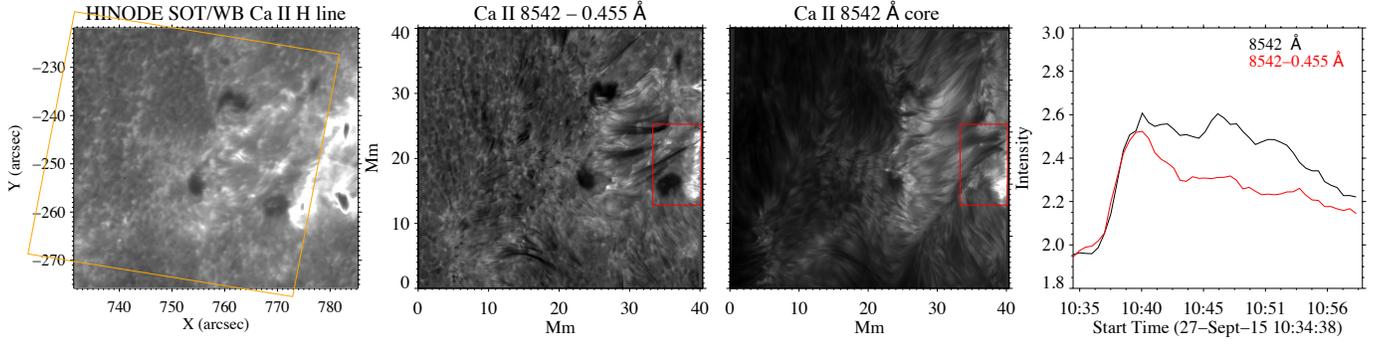}
\end{center}
\caption{The Ca {\sc{ii}} 8542 ${\AA}$ line wing and core images obtained with the CRISP instrument at 10:40:00 UT on 2015 September 27.  A context image of the flare in the Ca {\sc{ii}} H wide band 
obtained with the Solar Optical Telescope is shown on the left panel. The orange box indicates the flaring region observed with CRISP.  The temporal evolution of the region averaged over the 
20$\times$20 pixel$^2$ area on the 
flare ribbon is presented on the far right panel.}
\label{fig1}
\end{figure*}

\begin{figure*}[t]
\begin{center}
\includegraphics[width=17.9 cm]{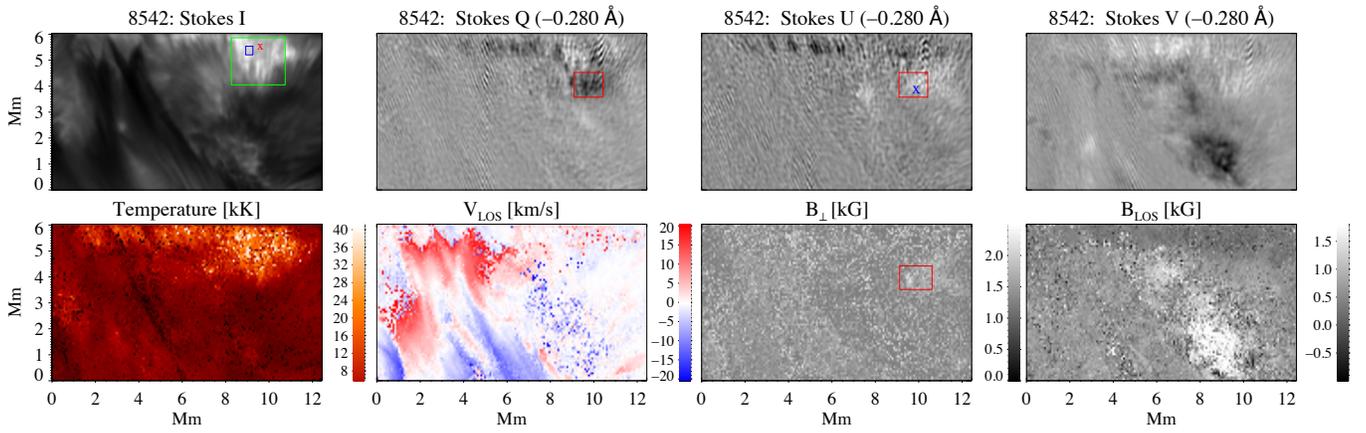}
\caption{The top row shows the SST images in Ca 8542 Stokes $I$ at line core and 
$Q,~U$ and $V$ at $\mathrm{\Delta\lambda = - 0.280~{\AA}}$  at 10:40:00 UT (flare peak) 
of the flaring region selected for inversions (marked with the red box in Figure~\ref{fig1}). 
The NICOLE output showing the temperature, LOS velocity, and LOS magnetic field maps
averaged over the interval between $\mathrm{log~\tau\approx-1.5~and-4}$,
and perpendicular (with respect to the observer) magnetic field map in the interval between $\mathrm{log~\tau\approx-1~and-2.5}$, are presented in the bottom panels.  The red and blue colors in the Dopplergrams represent positive Doppler velocities (downflows) and negative Doppler velocities (upflows), respectively.
The "x", the blue, green and red boxes mark the selected representative pixels and regions at the flare ribbon that are discussed in the text.}
\label{fig2}
\end{center}
\end{figure*}

The Ca {\sc{ii}} infrared (IR) triplet lines are well suited for flare diagnostics 
due to its good sensitivity to physical parameters, including magnetic field, in the solar photosphere and chromosphere  
\citep{2008A&A...480..515C,2007ApJ...663.1386P,2012A&A...543A..34D,2016MNRAS.459.3363Q}.
Furthermore, calcium is singly ionized under typical chromospheric conditions, with negligible non-equilibrium and partial redistribution effects for the Ca {\sc{ii}} IR lines \citep{1989A&A...213..360U,2011A&A...528A...1W}, 
making their modeling and interpretation of the observations more straightforward. Furthermore, the NLTE radiative transfer code NICOLE \citep{2015A&A...577A...7S}
%Furthermore, the optimization of NICOLE for Ca II 8542  allows the use of this feature for the computation of semi-empirical atmospheric models.
allows to perform the inversion of observed Ca {\sc{ii}} 8542 {\AA} (hereafter Ca 8542) Stokes profiles 
and the construction of semi-empirical model atmospheres. %{\bf ?? (one dimensional vertical stratifications of the physical parameters with 3D cubes produced by multiple such stratifications)??}. 
%NICOLE inversions of the Ca 8542 line allow for the ionization equilibrium, statistical equilibrium and polarized radiative transfer equations to be solved numerically to synthesize the Stokes profiles each time the atmospheric model is perturbed. 
%A $\chi^2$ minimization procedure uses the differences between the observed and synthetic profiles to iteratively modify, 
%at heights determined by the line response functions, the model atmospheres until the observed spectral signatures are reproduced.  
Such inversions have been successfully performed for spectropolarimetric Ca 8542 observations and 3D magnetohydrodynamical simulations of 
solar features in the umbra and penumbra of sunspots \citep{2013A&A...556A.115D,2017ApJ...845..102H}, granular-size magnetic 
elements (magnetic bubbles) in an active region, and magnetic flux tubes \citep{2015ApJ...810..145D,2017MNRAS.472..727Q}. 
Recently \cite{2017ApJ...846....9K} performed NICOLE inversions of high-resolution spectroscopic data in the Ca 8542 line and found that the temperature in the middle and 
upper chromosphere close to the flare peak is enhanced up to $\sim$6.5 - 20 kK between $\mathrm{log~\tau \sim-3.5~and-5.5}$, decreasing gradually to preflare temperatures of $\sim$5 - 10 kK  
approximately 15 minutes after the peak.

\begin{figure*}[t]
\begin{center}
\includegraphics[width=17.2 cm]{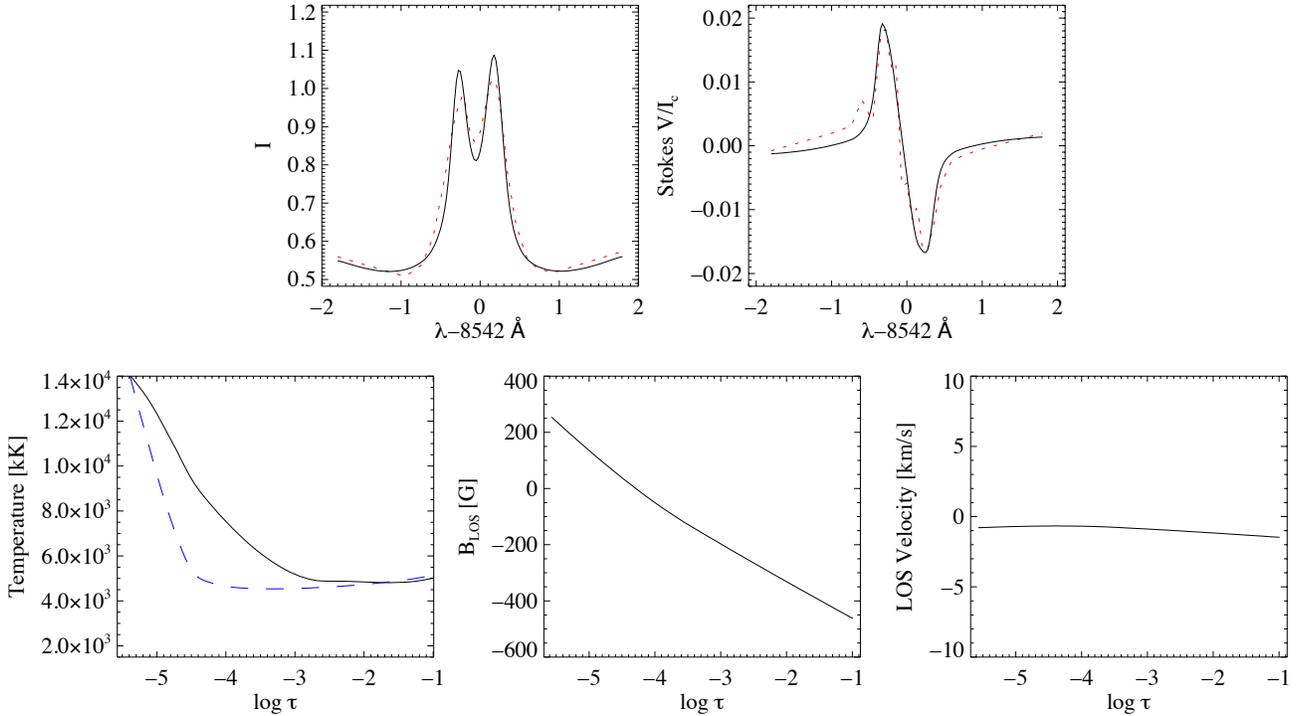}
\caption{A representative example of the observed (red dotted) and best-fit synthetic (black) Ca 8542 line Stokes $I$ and $V$ profiles together with temperature, velocity and 
LOS magnetic field stratifications for the flaring pixel 
marked with the red 'x' in Figure~\ref{fig2}. The blue dashed line depicts the typical temperature stratification obtained 
from inversions for a non-flaring pixel located at around x=1, y=4 Mm in the selected region (Figure 2).}
\label{fig4}
\end{center}
\end{figure*}

\begin{figure*}[t]
\begin{center}
\includegraphics[width=17.9 cm]{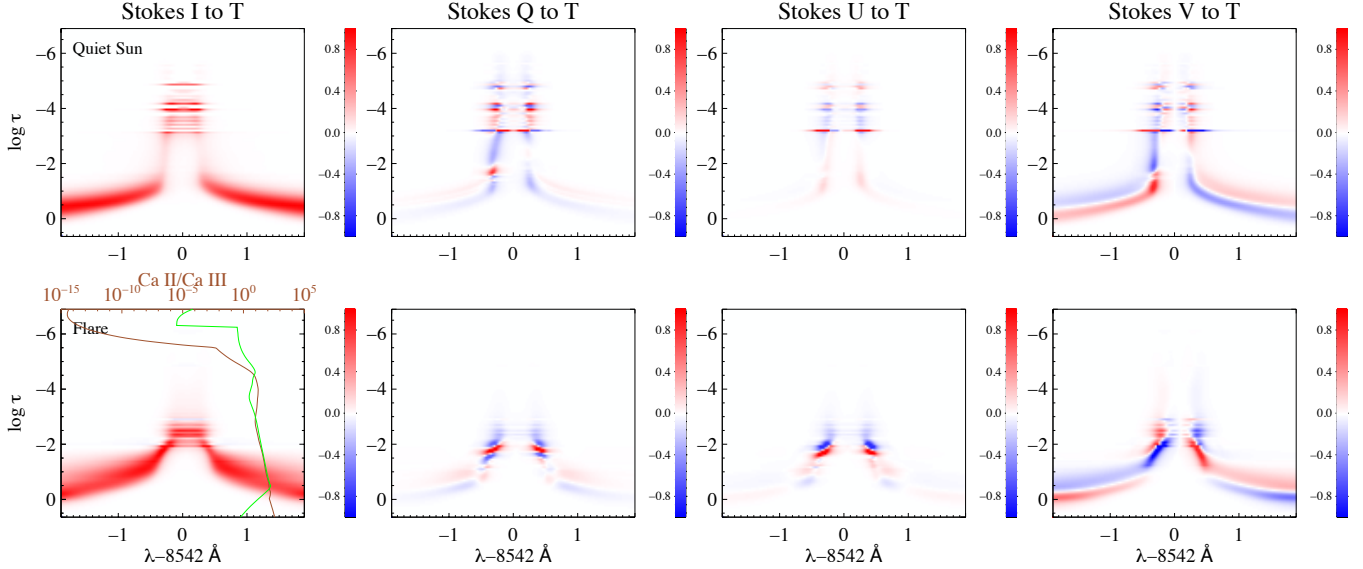}
\caption{Two dimensional plots of the RFs (normalised to their maximum values) which show how given Stokes profiles respond  
to changes in the temperature for Ca 8542 line for the area covering the non-flaring, quiet Sun region (top row) and the flare ribbon (bottom row) at 10:40:00 UT. 
The ratio of the population densities of Ca {\sc{ii}} and Ca {\sc{iii}} as a function of height for QS (green line) and flare (brown line) models are overplotted in the bottom left panel.}
\label{fig5}
\end{center}
\end{figure*}
\begin{figure*}[t]
\begin{center}
\includegraphics[width=17.9 cm]{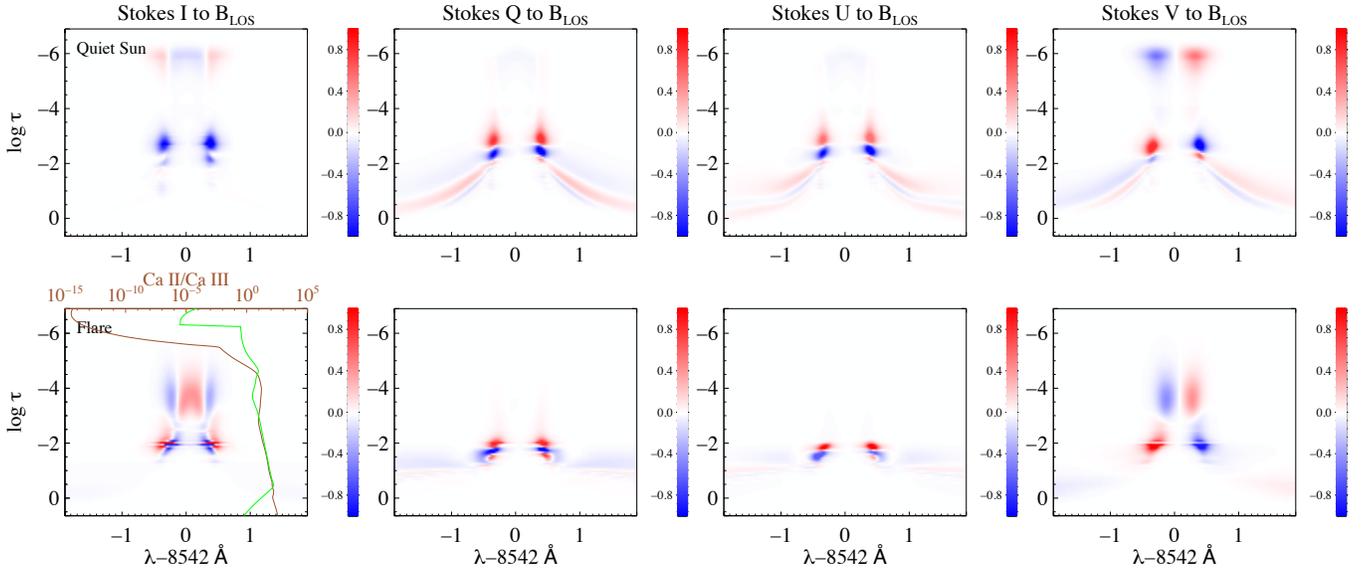}
\caption{Two dimensional plots of the RFs  which show how given Stokes profiles respond 
to changes in the LOS magnetic field for Ca 8542 line for the area covering the non-flaring, quiet Sun region (top row) and the flare ribbon (bottom row) at 10:40:00 UT. The ratio of the population densities of Ca {\sc{ii}} and Ca {\sc{iii}} as a function of height for QS (green line) and flare (brown line) models are overplotted in the bottom left panel.  
} 
\label{fig5}
\end{center}
\end{figure*}

Spectropolarimetric data of the flare in the Ca {\sc{ii}} 8542 and He {\sc{ii}} 10830 {\AA} lines have been analyzed in several works 
\citep{1995ApJ...441L..51P,2011A&A...526A..42S,2011SPD....42.0308K,2015ApJ...799L..25K,2015IAUS..305...73K,2014ApJ...796...85J,2015ApJ...814..100J}. 
Using photographic recordings of Stokes $I$ and $V$ profiles in the Ca {\sc{ii}} H line, \cite{1990ApJ...361L..81M} 
derived a value of 820~($\pm$40)~G for B$_{LOS}$ using the weak field approximation. \cite{2012SoPh..280...69H} studied the vector magnetic field in the Ca 8542 line and measured its LOS component for two flare kernels to be 415 and - 215 Gauss.
Recently, \cite{2017ApJ...834...26K} detected the stepwise chromospheric LOS magnetic field changes obtained through spectropolarimetry of Ca 8542 during an X-class flare. The Stokes $Q$ and $U$ signals across the line profiles and the associated transverse component of the magnetic field are more difficult to observe, as the variation of $Q$ and $U$ is, in general, more complex with less polarimetric sensitivity
than the Stokes $V$ profile and are more affected by instrumental smearing \citep{2017ApJ...848..107L}.
\cite{2012ApJ...748..138K} detected linear polarization of up to 10\% of the continuum level in the Ca 8542 line in the flare  ribbons located over the sunspot with typical symmetric Zeeman signatures.  

In this paper we analyze high-resolution imaging spectropolarimetric observations of an M-class solar flare in Ca 8542 line. We use NICOLE to invert the full Stokes profiles and construct models for the lower atmosphere of the flare event. Multiple inversions were performed for the observed region, covering flare ribbons and non-flaring areas at different times during the event.  The constructed models are used to obtain diagnostic information on key plasma parameters which focus on temperature, velocity, and magnetic field stratifications in the upper photosphere and chromosphere during the flare. The chromospheric magnetic field was also evaluated using the weak-field approximation method.

%%%%%%%%%%%%%%%%%%%%%%%%%%%%%%%%%%%%%%%%%%%%%%%%%%%%%%%%%%%%%%%%%%%%%%%%%%%%%%%%%%%%%%%%%%%%%%

\section{Observations and data reduction}
\label{sect:setup}

The observations were obtained between 10:34 and 10:59 UT on 2015 September 27 
close to the West limb ($774''$, $-217''$, $\mu\approx$ 0.545) with the CRisp Imaging SpectroPolarimeter \citep[CRISP;][]{2006A&A...447.1111S,2008ApJ...689L..69S} 
instrument, mounted on the Swedish 1-m Solar Telescope \citep[SST;][]{2003SPIE.4853..341S}
on La Palma. 
%Adaptive optics were used throughout the observations, consisting of a tip-tilt mirror 
%and in 85-electrode deformable mirror setup that is an upgrade of the system described in \cite{2003SPIE.4853..370S}. 
The observations comprised of imaging spectropolarimetry in Ca 8542 line. The data were reconstructed with Multi-Object Multi-Frame Blind Deconvolution \citep[MOMFBD;][]{2002SPIE.4792..146L,2005SoPh..228..191V}. 
We applied the CRISP data reduction pipeline as described in \cite{2015A&A...573A..40D} which includes small-scale seeing compensation for spectral scans as in \cite{2012A&A...548A.114H}.
Our FOV was $41\times41$\,Mm$^2$ with a spatial sampling was 0$''$.057 pixel$^{-1}$. The spatial resolution was close to the diffraction limit of the telescope at this wavelength (0$''$.18) for many images in the time-series.  
%The polarimetric calibration was performed using a linear polarizer and a quarter-wave {\bf plate?} at many different angles close to the primary focus on the optical table a few hours after the observations. A yearly calibration for the main lens and vacuum tube optics was also used as in \cite{2010arXiv1010.4142S}.
%CRISP records 4 LC states per wavelength which are a linear combination of the Stokes parameters. 
%These states are demodulated to obtain images of the full Stokes vector $I,~Q,~U$ and $V$ components using the calibration. 
All profiles were normalized to the continuum of the quiet-Sun intensity by fitting the FTS atlas profile \citep{1999SoPh..184..421N} convolved with the CRISP instrumental profile.

The Ca 8542 line scan consisted of 21 line positions with an irregular step size. From line core these were: -1.75 {\AA} to +1.75 {\AA} ($\pm$1.75,    $\pm$0.945,     
$\pm$0.735,     $\pm$0.595,    $\pm$0.455,     $\pm$0.35,     $\pm$0.28,    $\pm$0.21, $\pm$0.14,  $\pm$0.07,  0.0 {\AA}).
%Each spectral scan had an acquisition time of 15~s but the cadence 
The cadence of the time series was
22.3~s. 
The transmission FWHM for Ca 8542 is 107.3 m{\AA} with a pre-filter FWHM of 9.3 {\AA}.
Throughout the analysis we made use of CRISPEX \citep{2012ApJ...750...22V}, a versatile widget-based tool for effective viewing and exploration of multi-dimensional imaging spectroscopy data.

The M1.9 flare was observed on active region NOAA 12423. The rise phase and flare peak were captured during the observations. Figure~\ref{fig1} shows a sample of the flare images in the Ca 8542 line core and wing positions.
Unfortunately, our ground-based observations covered only $\sim20\%$ of the flare ribbon. Context imaging of the full ribbon are provided by the Ca {\sc{ii}} H wide band filtergrams 
obtained with the Solar Optical Telescope on board the HINODE satellite \citep{2007SoPh..243....3K,2008SoPh..249..167T} 
(see Figure~\ref{fig1}). Lightcurves generated from the region located on the ribbon show the evolution of the emission in the line core and wing positions (right panel of Figure~\ref{fig1}). 

%%%%%%%%%%%%%%%%%%%%%%%%%%%%%%%%%%%%%%%%%%%%%%%%%%%%%%%%%%%%%%%%%%%%%%%%%%%%%%%%%%%%%%%%%%%%%%

\section{Inversions}

\begin{table}[b]
\caption{Number of nodes and input atmosphere models used during each cycle (Cy) of the inversion.}
\begin{center}
\begin{tabular}{c   c   c   c  c}
\hline
Physical parameter ~&~~Cy 1~~&~~Cy 2& \\  
\hline    
Temperature    &        6 nodes        &         7  nodes       \\ 
LOS Velocity    &         2 nodes             &        3    nodes    \\ 
Microturbulence    &         1 node             &        1    node      \\ 
$\mathrm{B_z}$    &         2 nodes             &        3    nodes      \\
$\mathrm{B_x}$    &         2 nodes            &        3    nodes    \\
$\mathrm{B_y}$    &         2 nodes            &        3    nodes   \\
Macroturbulence    &         none             &          none    \\                 
Input atmosphere    &      Fal-C       &            model from Cy 1     \\
\hline
%\caption {}
\label{tabl1}
\end{tabular}
\end{center}
\end{table}

We used the NICOLE inversion code \citep{2015A&A...577A...7S} which
has been parallelized to invert large field of view (one pixel at a time) and solves multi-level, NLTE radiative transfer problems \citep{1997ApJ...490..383S}. 
The code iteratively perturbs physical parameters such as  temperature, LOS 
velocity, magnetic field and microturbulence 
of an initial guess model atmosphere to find the best match with the observations \citep{2000ApJ...530..977S}. 

We used a five bound level plus continuum Ca 8542 model atom 
with complete angle and frequency redistribution, which is applicable to lines such as Ca 8542  \citep{1974SoPh...39...49S,1989A&A...213..360U}. 
The synthetic spectra were calculated for a wavelength grid of 145 datapoints in 0.025 {\AA} steps, 
6 times denser than the CRISP dataset. Stratification of the atmospheric parameters obtained by the inversions are given as a function 
of the logarithm of the optical depth-scale at 500~nm (hereafter $\mathrm{log~\tau}$). 

To improve convergence, the inversions were performed in two cycles, expanding on the suggestion of \cite{1992ApJ...398..375R}. 
In the first cycle we use the FAL-C atmosphere \citep{1990ApJ...355..700F} as an initial model.
Table 1 summarizes the number of nodes and initial temperature models used in the two cycles.

We inverted a 300 $\times$ 160 pixel$^2$ area (12.3$\times$6.5 Mm$^2$) covering the flare ribbons as well as some non-flaring regions (see left panels of Figure~\ref{fig1}). The Stokes profiles were rebinned in $\mathrm{2\times2}$ pixels to increase the signal to noise ratio. Although our time-series comprises of 46 spectral scans, we only choose the best scans in terms of spatial resolution which limits the total number of inverted scans to 8.   

To investigate the sensitivity of the Ca 8542 line to changes of the atmospheric parameters at different heights in the lower solar atmosphere, 
we have also computed the numerical response function of the emergent profile  using the different models obtained from inversions.

%%%%%%%%%%%%%%%%%%%%%%%%%%%%%%%%%%%%%%%%%%%%%%%%%%%%%%%%%%%%%%%%%%%%%%%%%%%%%%%%%%%%%%%%%%%%%%

\begin{figure*}[t]
\begin{center}
\includegraphics[width=17.9 cm]{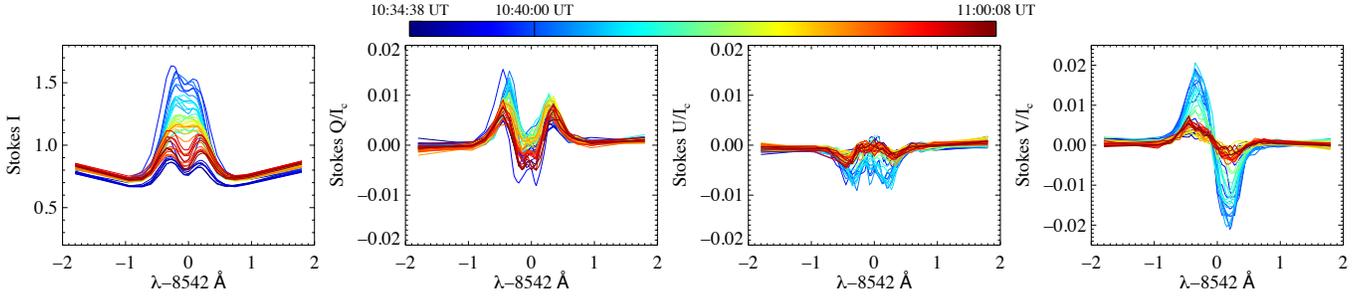}
\caption{Temporal evolution of the observed Ca 8542 Stokes profiles averaged over the region outlined with the blue box in Figure 2 starting $6$ minutes before the flare peak (10:40:00 UT).}
\label{fig5}
\end{center}
\end{figure*}

\begin{figure*}[t]
\begin{center}
\includegraphics[width=18.0 cm]{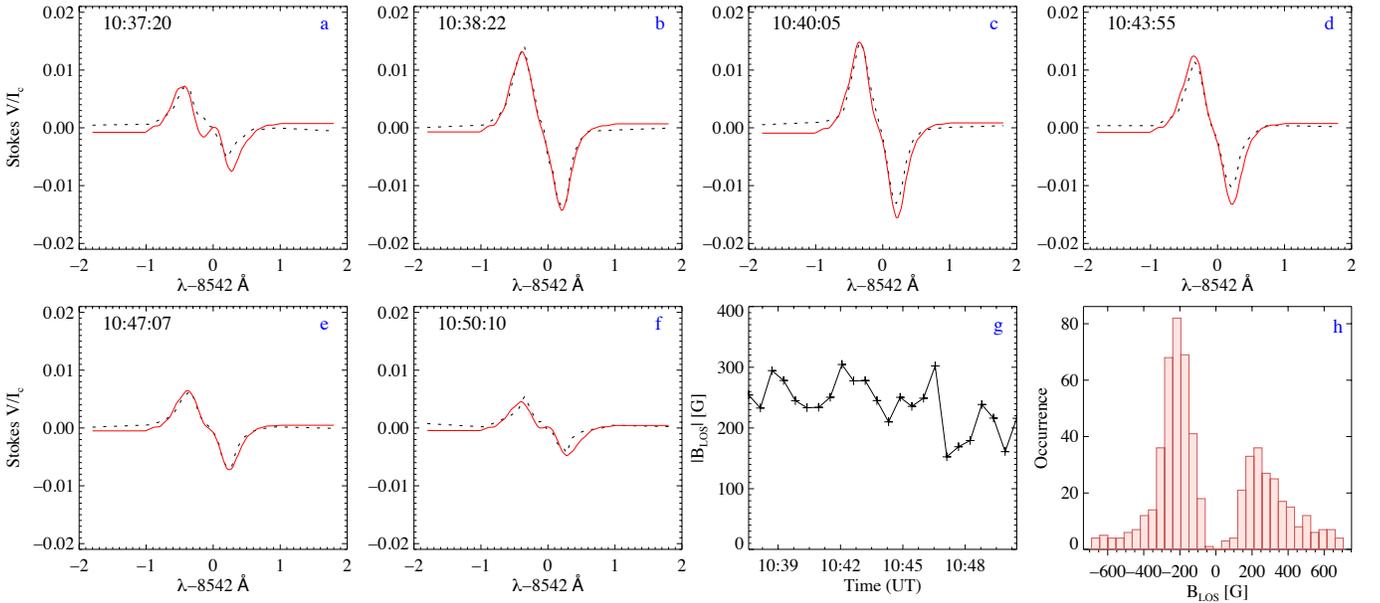}
\caption{(a-f) Circular polarization profiles (black dashed lines) for the flare ribbon at different phases of the event for the region outlined with the blue box in Figure 2. 
WFA fits obtained from the derivative of Stokes $I$ are depicted as full red lines. (g) The LOS magnetic field used to fit the Stokes $V$ as a function of time is presented in panel (g). 
(h) A histogram of the LOS magnetic field, from the fitting of Stokes $V$ profiles using the WFA for individual pixels of the area marked with the green box in Figure 2 during the flare peak (at 10:40:00 UT).} 
\label{fig6}
\end{center}
\end{figure*}

\section{Analysis and results}

The top row of Figure~\ref{fig2} presents the Ca 8542 Stokes $I$ at line core and 
$Q,~U$ and $V$ monochromatic images at $\mathrm{\Delta\lambda = - 0.280~{\AA}}$  at 10:40:00 UT (flare peak) 
of the region selected for inversions (marked with the red box in Figure~\ref{fig1}). 
The images show that the enhancement of circular and linear polarization signals (Stokes $V$ and $Q~\&~U$, respectively) are co-spatial with the flare ribbons observed in intensity (Stokes $I$).
The NICOLE output showing the temperature, LOS velocity, and LOS magnetic field maps
integrated between $\mathrm{log~\tau\approx-1.5~and-4}$ and averaged over this range, %{\bf integrated and averaged??}
and perpendicular ($B_\bot$ along LOS) magnetic field maps integrated and averaged between $\mathrm{log~\tau\approx-1~and-2.5}$,
are presented in the bottom panels of Figure~2.  A temperature map of the inverted region  
shows strong temperature enhancements for the ribbon. The LOS velocity map of the same region 
reveals weak upflows within the ribbon area and some strong downflows nearby where chromospheric fibrillar features are located (Figure~2). 
The inversion results for the perpendicular magnetic field map are very noisy and inconsistent throughout the FOV. 
However, these results trace the shape of a patch of strong (around 1 - 1.5 kG) perpendicular field along LOS for the part of the flare ribbon where linear polarization is strongest (area marked with a red box on Figure 2 and region immediately to the right displaying a wing-like shape). 
The LOS magnetic field map is less noisy and tracing magnetic field 
concentrations at the vicinity of a small sunspot located at the bottom right part of the FOV and at the flare ribbon (bottom right panel of Figure~\ref{fig2}).  

Figure~3 shows 
the observed Stokes profiles (red dotted lines) of the flaring pixel marked with red 'x' in Figure~\ref{fig2} at the flare peak ($\sim$10:40:00 UT). 
The best-fit synthetic profiles obtained from the inversion are also shown as full black lines. 
The temperature, LOS magnetic field and velocity stratifications reproducing the synthetic profiles of this pixel are presented at the bottom panels of Figure 3.
The Stokes $I$ profile is centrally reversed and shows excess emission in the red wing (red asymmetry) with a blueshifted line center (top left panel of Figure 3).  
The asymmetry appears to be due to the blueshifted line core of Stokes $I$ which in turn is a consequence of upflows (visible in our inversions in the bottom right panel of Figure 3) at the height of line core formation 
\citep{1997ApJ...481..500C,1999ApJ...521..906A,2015ApJ...813..125K,2016ApJ...832..147K,2017ApJ...846....9K}.
The flare pixel has a higher chromospheric temperature at $\mathrm{log~\tau\sim-2.5~and-5.5}$ compared 
to the quiet Sun temperature profile which is overplotted as a blue dashed line at the bottom left panel of Figure 3. The Stokes $Q$ and $U$ profiles
are very noisy which caused NICOLE to fail in the fitting of a perpendicular magnetic field stratification. 
The analysis of the line response functions shows that the inversions provide diagnostics for the LOS magnetic field in the layers between $\mathrm{log~\tau \sim-1~and-4.5}$ (see Figure 4 \& 5).
The absolute value of the LOS magnetic field in the lower atmosphere between $\mathrm{log~\tau\sim-1~and-4.5}$ decreases gradually with an almost linear trend from $\mathrm{|B_{LOS}|}$=450 G to $\sim$0 G (Figure 3).  

It must be noted that the imaging spectropolarimetric data analysed in this work are limited by the acquisition time required for  
a spectral scan of line profiles. 
Recently, \cite{2018arXiv180205028F} investigated the impact of the time-dependent acquisition on the inversion of spectropolarimetric profiles of Ca 8542 line during umbral flashes.
They showed that the inversion results could be unreliable at the development phase of the flash when the atmosphere is changing rapidly due to the time-dependent acquisition of line profiles.
They have considered an extreme case where the line profiles are changing from absorption to emission during the scan.
The flaring atmosphere can also change rapidly during the scanning time, especially at the impulsive phase of the flare which is characterised by very short dynamical timescale. 
However, temporal evolution of line profiles analysed in this work are not suggesting that they are changing significantly during the scan.
Furthermore,  lightcurves generated from the region located on the ribbon also indicates that flare has a much longer evolution timescale when compared to the acquisition time of the Ca 8542 line profiles ($\mathrm{\sim15~s}$) (see right panel of Figure~1).

To investigate how sensitive the Ca 8542 line is to the different layers of the atmosphere (in $\mathrm{log~\tau}$ units) covered by the models, we need to examine the response function (RF), which measures the response of the emergent profiles to perturbations of the physical parameters.
We compute the numerical RFs for the Stokes profiles for different atmospheric models obtained from NICOLE inversions. 
Figures 4 and 5 show 2 dimensional plots of the RFs of the given Stokes profiles 
to changes in the temperature and LOS magnetic field for the area covering the flare ribbon and for the non-flaring quiet Sun region. 
RFs to the temperature of the QS Ca 8542 Stokes profiles, covering a $\mathrm{\pm1.75~{\AA}}$ spectral range from the line core,  
are expanded to $\mathrm{log~\tau\sim -1~and-5}$. However, Stokes $Q~\&~U$ have very small sensitivity 
to temperature in the outer wings beyond $\mathrm{\pm1~{\AA}}$ from the line core (top panels of Figure 4). 
For the flare model RFs to temperature for the Stokes profiles have enhancements between $\mathrm{log~\tau\sim -1~and -3}$ with very small values  between $\mathrm{log~\tau\sim -3~and -4.5}$. 

The RFs to the LOS magnetic field of the QS model for the Ca 8542 Stokes profiles show responses up to $\mathrm{log~\tau\sim -6}$
with a strong peak between $\mathrm{log~\tau\sim -1.5~and \sim-3.5}$ (top panels of Figure~5).
However, the peaks of the RFs to the LOS magnetic field of the flare models are located deeper in the atmosphere between $\mathrm{log~\tau\sim -1~and -2.5}$  (bottom panels of Figure~5).
This indicates that the Ca 8542 Stokes profiles for the flare models are more sensitive to perturbations of the temperature and magnetic field in the deeper atmospheric layers than for QS atmosphere models. 
The RFs computed for multiple pixels from different regions consistently show similar changes in the sensitivity of Ca 8542 throughout the different QS and Flare models.
To investigate the reason for this difference we analyzed the  ionization level of Ca 8542 line in QS and flare models by computing the population densities of all Ca {\sc{ii}} atomic levels over the population densities of Ca {\sc{iii}}. The results for the QS (green line) and the flare (brown line) models are  
overplotted on the bottom left panels of Figure 4 and 5. The results show that the fraction of Ca {\sc{ii}} to Ca {\sc{iii}} decrease 
by about $\sim$100 times above $\mathrm{log~\tau\sim -4.5}$ when compared to the Ca {\sc{ii}}/Ca {\sc{iii}} in the QS atmosphere (Figure 4 \& 5) indicating that Calcium is almost fully ionized to Ca {\sc{iii}} above
$\mathrm{log~\tau\sim -4.5}$. This can explain why the Ca 8542 Stokes profiles are less sensitive to the physical properties of the upper atmosphere during the flare.

Figure 6 displays the evolution of the Ca 8542 Stokes profiles located on the flare ribbon and averaged over the $\mathrm{\sim0.35''\times0.35''}$ area marked with the blue box in Figure 2. It shows that, as the flare initiates, the Stokes profiles of the area located on the ribbon increase in intensity, becoming strongest during the flare maximum and 
decreasing after the maximum. 
An enhancement of polarization signals could be related to the increase of Stokes $I$ amplitude. 
Stokes $V,~Q,~\&~U$  
depend on the first and second derivatives of the intensity (Stokes $I$) in the weak field approximation (WFA) regime. 
We can apply the WFA here as the field strength is expected to be significantly lower than 2.5 kG and the Zeeman splitting is much smaller than the thermal width of the observed Ca 8542 line \citep{2013A&A...556A.115D}.
In this regime the Stokes profiles can be 
expressed as \citep{1977A&A....56..111L},   
\begin{equation}
V(\lambda)= - 4.67\times10^{-13}g_{eff}\lambda_0^2B_{LOS}\frac{\partial I(\lambda)}{\partial \lambda},
\end{equation}
\begin{equation}
\left[Q^2+U^2\right]^{1/2}= \left |- 5.45\times10^{-26}\bar{G}\lambda_0^4B^2_{\bot}\frac{\partial^2 I(\lambda)}{\partial \lambda^2}\right |,
\end{equation}
where $g_{eff}$ is the first order effective Land\'e factor,  
and $\lambda_0$ the central wavelength of the spectral line.
$\bar{G}$ is a second order effective Land\'e factor related to $g_{eff}$ with
\begin{equation}
\bar{G}=g_{eff}^2-(g_1-g_2)^2(16s-7d^2-4)/80,
\end{equation}
where
$$
s=J_1(J_1+1)+J_2(J_2+1),~~d=J_1(J_1+1)-J_2(J_2+1)
$$
for the angular momentum $J_1$ and $J_2$ of the involved energy levels with Land\'e factors  $g_1$ and $g_2$ \citep{1977A&A....56..111L}.
For Ca 8542 $g_{eff}$=1.1 and $\bar{G}$ =1.21. We note that  the units for the
wavelength and magnetic field are {\AA} and Gauss, respectively.   

Figure 7 (panels a - f) shows average Stokes $V$ profiles for the region outlined with the blue box in Figure 2 
during the different phases of the flare together with the WFA fits obtained from the derivative of Stokes $I$ (overplotted as full red lines). The temporal evolution of the LOS magnetic field ($|B_{LOS}|$) 
used to fit the Stokes $V$ profiles at different times is presented in panel (g) of Figure~7.  It shows that the average LOS magnetic field obtained with the 
WFA is around 250 $\pm$ 50 G until $\sim$10:47 UT, decreasing later down to 180 $\pm$ 30 G. Increased gradients in Stokes $I$, as those observed in the flare, 
lead to increased polarization signals and this is described in the WFA (Equations 1 \& 2). Gradients of Stokes $I$ with respect to wavelength then certainly contribute to the observed enhanced 
amplitude of polarization signals (as in de la Cruz Rodr\'{i}guez et al. 2012, Quintero Noda et al. 2017). However, 
such increases did not lead to WFA reproducing the flaring profiles without an increase in magnetic field. 
The increase in magnetic field could    
be related to the change of sensitivity of Ca 8542 in the chromosphere during the flare to lower layers where the field is expected to be stronger. 
As the atmosphere relaxes after the flare peak, the line becomes more sensitive to the higher levels of the atmosphere (Figure 4 \& 5) where the magnetic field is expected to be weaker. 
A histogram of the LOS magnetic field computed with the fitting of Stokes $V$ using the WFA for individual pixels of the area marked with the green boxes in Figure 2 during the flare peak, is presented in panel (h) of Figure 7 showing median absolute values of 220$\pm$70 G.

\begin{figure}[t]
\begin{center}
\includegraphics[width=8.7 cm]{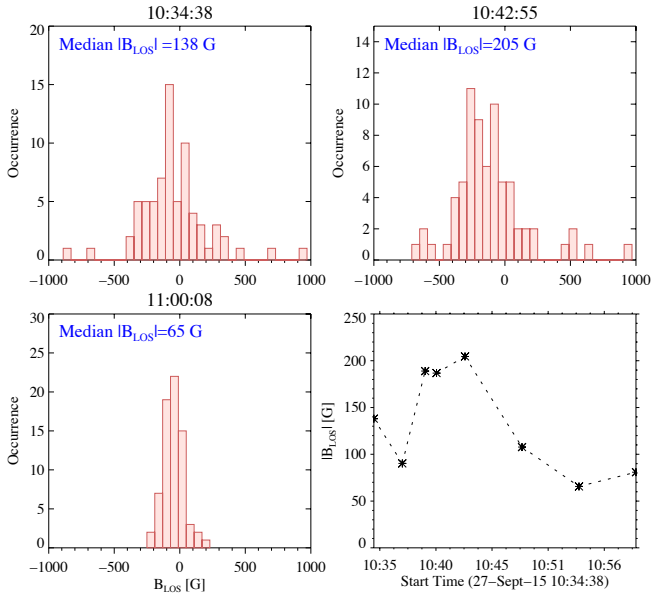}
\caption{Histograms of $|B_{LOS}|$ 
computed from the semi-empirical models obtained from the inversions before, during and after the flare peak
for the pixels of the area delimited with the green box in Figure 2 (after filtering for quality of fit). $|B_{LOS}|$  is averaged over optical depths ranging from $\mathrm{log~\tau\sim -1.5~to -4.5}$. 
Medians of $|B_{LOS}|$ for the same area obtained at 8 different points in time are presented in the bottom right panel.} 
\label{fig6}
\end{center}
\end{figure}

Figure~8 shows the histograms of $|B_{LOS}|$ 
computed from the inversions before, during and after the flare peak
for individual pixels of the area marked with the green box in Figure 2. $|B_{LOS}|$  is averaged over optical depths $\mathrm{log~\tau\sim -1.5~to -4.5}$. 
Models obtained at 8 different points in time were used to compute the median of $|B_{LOS}|$ for the same area. 
This temporal evolution is presented at the bottom right panel of Figure 8. It shows that 6 minutes before the flare peak the median of the integrated $|B_{LOS}|$ for the region marked with the green box in Figure 2 
is around 138 G, increasing to 205 G at the flare peak (bottom right panel of Figure 8). After 10:47 UT $|B_{LOS}|$  decreases gradually to $\sim$70 G. We note that the models of the pixels with a low quality of fit were ignored and are not included in the histogram plots. The inverted LOS magnetic field evolution close to the flare peak, between 10:38 and 10:45 UT, is consistent with that of the $B_{LOS}$ evolution obtained with the WFA  (shown in panel (g) of Figure 7), in that  they show similar values and a decrease post 10:45 UT. 

Compared with the LOS component, it is more challenging to compute the perpendicular component of the magnetic field along LOS with the WFA fitting 
as it is related to the total linear polarization, which depends on the second derivative of Stokes $I$ with respect to wavelength (Equation 2).   
Unfortunately, the observed Stokes $Q~\&~U$ can not be fitted reliably with the second derivative of Stokes $I$ under the WFA limit (Equation 2).
To improve the quality of the signal we identified the area of the flare ribbon with the strongest amplitude in Stokes $Q~\&~U$ (red box on Figure 2),
and computed averaged profiles over that $0.23''\times0.23''$ area, followed by a NICOLE inversion. The representative example of Stokes profiles with the best-fit synthetic profiles obtained from the inversion 
for the pixel marked with the blue 'x' in the red box on Figure 2 are presented on Figure 9. Temperature, LOS, and perpendicular magnetic field stratifications 
are shown on the bottom panels of Figure 9. The analysis of the response functions to the $B_{LOS}$ of Ca 8542 during the flare shows 
that inversions can provide diagnostics in the layers between $\mathrm{log~\tau\sim -1~and -2.5}$, 
with very small sensitivity between $\mathrm{log~\tau\sim -2.5~and -4}$  for Stokes $Q~\&~U$, and  
$\mathrm{log~\tau\sim -1~and -4}$ for Stokes $V$. $|B_{LOS}|$ in the atmosphere between $\mathrm{log~\tau\sim-1~and-4}$ is about $\sim$400-200 G whereas the $B_{\bot}$ 
in the atmosphere between $\mathrm{log~\tau\sim-1~and-2.5}$ is about 1.4 KG.

Despite the improved polarization signals, the total linear polarization ($\sqrt{Q^2+U^2}$) still can not be reliably fitted with the second derivative of Stokes $I$ under the WFA limit (Equation 2). 
Applying the WFA to the synthetic, best-fit profiles obtained from the inversion, led to a resulting total linear polarizatio that does not completely trace with the synthetic one (bottom left panel of Figure 9), however the resulting perpendicular component of the magnetic field 
is around 1300 G which is very close to the values obtained from the inversion (bottom third panel of Figure 9).

%%%%%%%%%%%%%%%%%%%%%%%%%%%%%%%%%%%%%%%%%%%%%%%%%%%%%%%%%%%%%%%%%%%%%%%%%%%%%%%%%%%%%%%%%%%%

\section{Discussion and conclusions}

\begin{figure*}[t]
\begin{center}
\includegraphics[width=17.9 cm]{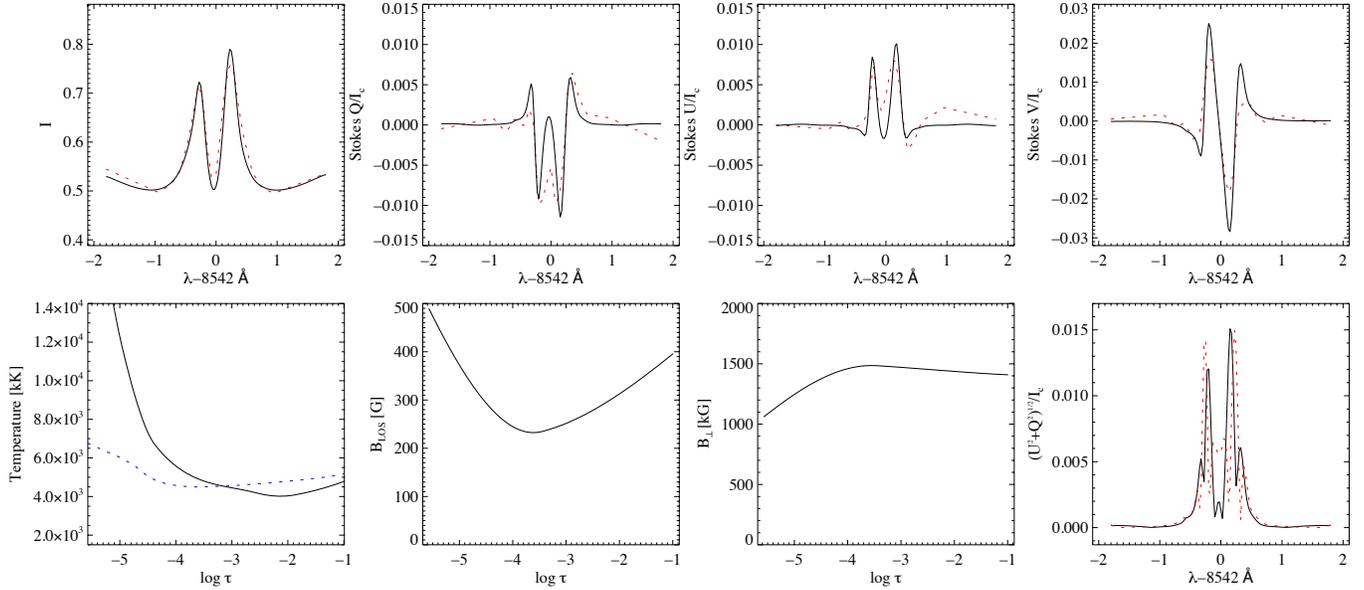}
\caption{A typical set of Stokes profiles with corresponding well fitting synthetic profiles obtained from the inversion 
of a pixel from within the red box in Figure 2 averaged over that $0.23''\times0.23''$ area, selected due to Stokes $Q$ and $U$ being successfully fitted. Temperature, LOS, and perpendicular magnetic field stratifications are shown in the bottom panels.
The bottom right panel shows the synthetic, total linear polarization, $\sqrt{Q^2+U^2}$ (full black lines) and WFA fit (red dotted line) obtained from the second derivative of synthetic Stokes $I$.}
\label{fig5}
\end{center}
\end{figure*}

We studied an M1.9 class solar flare using high-resolution full-Stokes imaging spectropolarimetric observations in the Ca {\sc{ii}} 8542\,\AA\  line. We used the non-LTE code NICOLE to construct semi-empirical models of the lower flaring atmosphere and investigate the temperature structure and magnetic field of the flare. Our analysis shows that the most intensively heated layers in the flaring lower atmosphere are the middle and upper chromosphere between optical depths  $\mathrm{log~\tau\sim -2.5~and -5}$, with temperatures between $\sim$5 and 13 kK (Figure 2 and 3). In the photosphere, below $\mathrm{log~\tau\sim -2.5}$, there is no significant difference in the temperature stratification between quiescent and flaring atmospheres (Figure 3). 
This agrees with the previous results obtained by \cite{2017ApJ...846....9K} showing that during the C8.3 flare the atmosphere remained unchanged below $\mathrm{log~\tau\sim -2.5}$. 

Our analysis shows that during flare maximum, the polarization signals of the ribbon in the chromosphere become stronger compared to its surroundings. To investigate if this relates to a change in the line formation height, we computed numerical RFs for the different semi-empirical models. The comparison of the RFs to perturbations in temperature and longitudinal magnetic field, between flaring and non-flaring atmospheres, shows that in the flaring atmosphere the Ca 8542 line is more sensitive to the lower layers than in the quiet Sun. This suggests that 
the formation of the line in the flaring atmosphere occurs in the deeper layers where the magnetic field is expected to be stronger which may result in the enhanced polarization signals.
This analysis also indicates that the rapid flare heating of the chromosphere,  
significantly changes the balance between the population densities of the energy states in the Calcium atom and, most importantly, the fraction between Ca {\sc{ii}} and Ca {\sc{iii}} (Figure 4 $\&$ 5). 
The latter decreases 
by a factor of 100 above $\mathrm{log~\tau\sim -4.5}$ when compared to Ca {\sc{ii}}/Ca {\sc{iii}} in the QS atmosphere indicating that  Ca {\sc{ii}} is almost fully ionized to Ca {\sc{iii}} above
$\mathrm{log~\tau\sim -4.5}$. This can explain why the Ca {\sc{ii}} Stokes profiles are less sensitive in the higher layers of the atmosphere during the flare (Figure 4 $\&$ 5). This is in agreement with \cite{2016ApJ...827..101K} who show that, in dynamic flare models, the Ca {\sc{ii}} becomes fully ionized above $\sim$600 km due to intense heating.

The averaged LOS magnetic field of the flare ribbon, as obtained from inversions  
over the range of $\mathrm{log~\tau\sim -1~to -4.5}$, is estimated as 150~G 6 minutes before flare maximum, 
increasing to $\sim$200~G at the flare peak, and decreasing to $\sim$65~G approximately 7 minutes after the peak (Figure 8). 
The chromospheric magnetic field was also determined using the WFA, which led to a very similar value of 220$\pm$70 G for the flare peak. The WFA approach also led to a post-flare decrease, although this was less pronounced compared to the inversion outputs ($\sim$170 $\pm$ 30 G approximately 6 minutes after the peak).

We note that for most pixels in the detailed analysis area (green box in Figure 2), 
there are very good matches between Stokes $V$ and $\partial I/\partial \lambda$ suggesting that the increase of Stokes $I$ amplitude is the main reason for the enhanced amplitude of polarization signals during the flare. %Such increase in Stokes I can also come 

Estimating the transverse/perpendicular component of the magnetic field is extremely challenging 
as the linear polarization has low sensitivity for weak fields and the profiles obtained are usually noisier than those of circular polarization. Our inversions have led to good fits in a patch of strong 
 $\mathrm{\sim1-1.5}$ kG (fitted) perpendicular field, for the portion of the flare ribbon where linear polarization is the strongest (areas marked with red boxes on Figure 2). This is an unusually strong magnetic field for the chromosphere. We also applied the WFA fitting to the synthetic total linear polarization profiles obtained from the inversions and were able to reproduce the observed linear polarization profiles (Figure 9). The resulting perpendicular (with respect to the LOS) component of the magnetic field inside the area marked with the red box on Figure 2 is around 1300 G which is very close to the value obtained from inversions (bottom third panel of Figure 9).

LOS magnetic fields of the order of kG for chromospheric flare ribbons  have been reported in previous studies. For example, \cite{1990ApJ...361L..81M} derived a value for B$_{LOS}$ of 820 ($\pm$40) G with Stokes $I$ and $V$ 
profiles from the Ca {\sc{ii}} H~line and the WFA. \cite{2012SoPh..280...69H} evaluated B$_{LOS}$ for a flare emission kernel to be 415~G using the Ca {\sc{ii}} 8542 {\AA} line, also with the WFA.
\cite{1995ApJ...441L..51P} used the 10830 {\AA} He {\sc{i}} line and found mean B$_{LOS}$ of about 735~G during the decay phase of the flare kernel. However, to our knowledge, perpendicular components of the magnetic field with respect to the observer of the order of kG have never been observed in the flaring chromosphere. 
Yet, if the perpendicular magnetic field with respect to the surface of the Sun close to the disk center can reach up to kG level in the flaring chromosphere then, 
away from the disk center, this field can be observed as the strong perpendicular field with respect to the LOS. 
The observed region analyzed in this work is located away from the disk center 
(with $\mu\approx$ 0.545) and hence the magnetic field of the flaring loops located in this particular area of flare ribbon (red box on Figure 2)
can be oriented perpendicularly with respect to the observer. This can be a reason of an observed enhanced 
amplitudes of linear polarization signal and strong perpendicular component of the magnetic field.

Our analysis shows that the increased amplitude of the Stokes profiles including polarization signals related to flare energy release in the lower solar atmosphere 
allow reliable measurement of the chromospheric magnetic field. Good agreement between the measured magnetic field using
multi-node inversions in NLTE and WFA fitting, strongly suggests that both methods can be successfully applied to the flaring chromosphere. %{\bf last sentence is somewhat repetitive?}

%%%%%%%%%%%%%%%%%%%%%%%%%%%%%%%%%%%%%%%%%%%%%%%%%%%%%%%%%%%%%%%%%%%%%%%%%

\begin{acknowledgements}

The research leading to these results has received funding from the 
S\^er Cymru II Part-funded by the European Regional Development Fund through the Welsh Government.
The work of DK was supported by Georgian Shota Rustaveli National Science Foundation project FR17\_323.
DK and MM acknowledge the support provided by an STFC Consolidated Grant to Queen's University Belfast. 
The Swedish 1-m Solar Telescope is operated on the island of La Palma 
by the Institute for Solar Physics (ISP) of Stockholm University at the Spanish Observatorio del Roque de los Muchachos of the Instituto de Astrof\'{\i}sica de Canarias.
The SST observations were taken within the Transnational Access and
Service Programme: High Resolution Solar Physics Network (EU-7FP
312495 SOLARNET). 
Hinode is a Japanese mission developed by ISAS/JAXA, with the NAOJ as domestic partner and NASA and STFC (UK) as international partners. It is operated in cooperation with ESA and NSC (Norway). 
VMJH and LRvdV  were supported by the Research Council of Norway (project 250810/F20) and, together with MC, also through its Centres of Excellence scheme, project number 262622.
This project has received funding from the European Research Council (ERC) under the European Union's Horizon 2020 research and innovation programme (SUNMAG, grant agreement 759548). JdlCR is supported by grants from the Swedish Research Council (2015-03994), the Swedish National Space Board (128/15) and the Swedish Civil Contingencies Agency (MSB).

\end{acknowledgements}

\bibliography{bibtex.bib}
\end{document}